\UseRawInputEncoding
\documentclass[preprint,12pt,times]{elsarticle}

\usepackage{amssymb}

\usepackage{hyperref}
\usepackage{graphicx}
\usepackage{subcaption}
\usepackage{booktabs} 
\usepackage{multirow}
\usepackage{mathtools}
\usepackage{lineno}

\journal{arxiv}

\begin{document}

\begin{frontmatter}

%% Title, authors and addresses

%% use the tnoteref command within \title for footnotes;
%% use the tnotetext command for theassociated footnote;
%% use the fnref command within \author or \address for footnotes;
%% use the fntext command for theassociated footnote;
%% use the corref command within \author for corresponding author footnotes;
%% use the cortext command for theassociated footnote;
%% use the ead command for the email address,
%% and the form \ead[url] for the home page:
% \title{Title\tnoteref{label1}}
% \tnotetext[label1]{}
% \author{Name\corref{cor1}\fnref{label2}}
% \ead{email address}
% \ead[url]{home page}
% \fntext[label2]{}
% \cortext[cor1]{}
% \affiliation{organization={},
%             addressline={},
%             city={},
%             postcode={},
%             state={},
%             country={}}
% \fntext[label3]{}

\title{Design and Characterization of a Novel Motion Conversion Element: Curved Groove Ball Bearing without Retainer}

%% use optional labels to link authors explicitly to addresses:
\author[nwafu]{Shaoxiang Wang}

\author[nwafu]{Kaili Zhang}
\author[ZN]{Yaohua Hu}

\author{Lixia Hou\corref{cor1}\fnref{nwafu}}
\ead{houlixia@nwsuaf.edu.cn}
\cortext[cor1]{corresponding author}

% \author[nwafu]{Kangquan Guo}
\author{Kangquan Guo\corref{cor2}\fnref{nwafu}}
\ead{jdgkq@nwsuaf.edu.cn}
\cortext[cor2]{corresponding author}
% \fntext[fn2]{College of Mechanical and Electronic Engineering,Northwest A\&F University}

\affiliation[nwafu]{organization={College of Mechanical and Electronic Engineering},
            addressline={Northwest A\&F University},
            city={Yangling},
            postcode={712100},
            state={Shaanxi},
            country={China}}
            
\affiliation[ZN]{organization={College of Optical, Mechanical and Electrical Engineering},
            addressline={Zhejiang A\&F University},
            city={Hangzhou},
            postcode={311300},
            state={Zhejiang},
            country={China}}

\begin{abstract}

% An innovative mechanical transmission component, the curved groove ball bearing without retainer (CGBBR), 
% is proposed in this paper to facilitate the conversion between rotary motion and reciprocating motion.
% Compared to traditional mechanisms for motion conversion, CGBBR has several advantages, 
% such as a simple and compact structure and the absence of high-order vibrations. 
% This paper describes the design methodology, structural characteristics, kinematic laws, 
% and dynamic response characteristics of CGBBR, as well as its influencing factors. 
% A design method combining the curve family envelope theory and swept surface 
% theory is presented to create the unique spatial surface structure of CGBBR. A design scheme for 
% CGBBR has been presented in this paper, which adopts the concept of diameter-stroke ratio. 
% Additionally, a calculation method has been proposed for determining the number of rolling 
% elements in CGBBR. This method serves as the basis for subsequent optimization design. 
% The kinematic laws and vibration characteristics of CGBBR are analyzed, which ultimately 
% demonstrate its application potential, offering new insights for the design of mechanical 
% devices such as engines and compressors.
An unconventional and innovative mechanical transmission component, the Curved Groove Ball Bearing 
without Retainer (CGBBR), is introduced in this paper to facilitate the conversion between rotary 
and reciprocating motion. The CGBBR boasts several advantages over conventional motion conversion 
mechanisms, including a streamlined and compact structure, as well as the mitigation of high-order 
vibrations. This paper delves into the design methodology, structural attributes, kinematic principles, 
and dynamic response properties of the CGBBR. A novel design method named closed curve envelopment 
theory is proposed to generate the distinctive spatial surface structure of the CGBBR. Moreover, 
a design scheme is presented that employs the concept of diameter-stroke ratio for CGBBR implementation. 
A calculation methodology is also introduced for determining the number of rolling elements within the 
CGBBR, which serves as the foundation for subsequent optimization design. This paper deeply analyzes 
the kinematics principle of CGBBR and provides a new insight of motion conversion in mechanism.

\end{abstract}

% %Graphical abstract
% \begin{graphicalabstract}
% \includegraphics[width=1\textwidth]{images/bearing_move.pdf}
% This graphic abstract showcases the innovative design and application potential of curved groove ball bearings without a retainer (CGBBR) for efficient rotary-to-reciprocating motion conversion. CGBBR's compact structure and unique design, based on diameter-stroke ratio, are developed using envelope theory and swept surface theory. The proposed calculation method for the number of rolling elements paves the way for optimization. The analysis of kinematic laws and vibration characteristics demonstrates CGBBR's promise for mechanical devices, such as engines and compressors, offering a fresh perspective on mechanical transmission component design.
% \end{graphicalabstract}

%%Research highlights
% \begin{highlights}
% \item CGBBR, a novel motion conversion mechanism, offers a compact structure and eliminates vibration issues in traditional mechanisms.
% \item Envelope theory and swept surface theory applied to design CGBBR, successfully converting rotary and reciprocating motions.
% \item Diameter-stroke ratio and rolling element calculation method proposed for CGBBR optimization and design.
% \item Unique periodic pulse cluster phenomenon in CGBBR vibration signals identified and investigated.
% \item The potential applications of CGBBR include engines, compressors, and industrial robot joints, promising significant impact in various fields.
% \end{highlights}

\begin{keyword}
%% keywords here, in the form: keyword \sep keyword
Ball bearing \sep Motion conversion \sep Amplitude \sep Motion cycle
\sep Curved groove
%% PACS codes here, in the form: \PACS code \sep code

%% MSC codes here, in the form: \MSC code \sep code
%% or \MSC[2008] code \sep code (2000 is the default)

\end{keyword}

\end{frontmatter}

% \linenumbers

%% main text
\section{Introduction}
\label{Introduction}

Rotational motion and reciprocating motion are prevalent types of motion in engineering. They are commonly observed in the motions of the crankshaft and piston of engines and compressors, among others\cite{sclaterMechanismsMechanicalDevices2011}. In the 
field of mechanical design, there is a constant requirement for transformation between rotational motion and reciprocating motion. However, motion conversion mechanisms are 
restricted to a few select structures, such as the crankshaft, cam and toggle mechanisms\cite{uickerTheoryMachinesMechanisms2017}. 
These motion conversion mechanisms possess distinct advantages and drawbacks. Therefore, it is crucial to develop innovative motion conversion mechanisms in order to optimize the structural design of various mechanical devices, including engines, from both a theoretical and practical perspective.

The transformations between rotary and reciprocating motion are frequently achieved through the use of crank-slider and ball screw. The study of crank-slider and ball screw is well-established in the 
field of engineering. The available literature focuses on the structural design
\cite{zhouAdjustableSlidercrankLinkages2002,yaminPerformanceSimulationFourstroke2004a,soylemezClassicalTransmissionangleProblem2002,ozcanPerformanceEmissionCharacteristics2008}, 
vibration characteristics\cite{boysalTorsionalVibrationAnalysis1997,geveciInvestigationCrankshaftOscillations2005,chenKinematicAccuracyNonlinear2021,mourelatosCrankshaftSystemModel2001,erkayaDynamicAnalysisSlider2007}, 
wear\cite{pontNumericalDynamicAnalysis2017} and failure causes
\cite{fonteFailureAnalysisDamaged2019,liAnalysisCrankshaftFatigue2015,witekStressFailureAnalysis2017} of the crank-slider, as well as the 
transmission accuracy\cite{wenDegradationAssessmentBall2018,changFractalAnalysisVibrational2013,miuraDesignHighprecisionBall2017}, forward and reverse transmission stability\cite{weiKinematicalAnalysesTransmission2011,fengInvestigationBallScrew2012} of the ball screw. Crank-slider 
inherently possess high-order inertial forces as a result of their asymmetrical structure, which 
causes vibration and noise in operation. Ball screw has a complex structure\cite{huKinematicsBallscrewMechanisms2014}, resulting in 
difficulty in installation and maintenance\cite{hanTechniquesDevelopedFault2018}. In contrast, curved groove ball bearing without retainer (CGBBR) is composed entirely of 
axisymmetric rotatable structures. Components of CGBBR exhibit either rotational or 
reciprocating linear motion, without space translation or any other complex forms of motion, 
which are difficult to balance\cite{guoCurvedGrooveBall2021,shiManufactureCarryingCapacity2017a}. Therefore, CGBBR can eliminate typical issues of vibration 
and noise in traditional mechanical structures effectively, providing a new solution to the 
design and optimization of mechanical structures\cite{guoCurvedGrooveBall2022,guoCurvedTrenchBall2018}.

\begin{figure}[htbp]
	\centering
	\includegraphics[width=1\textwidth]{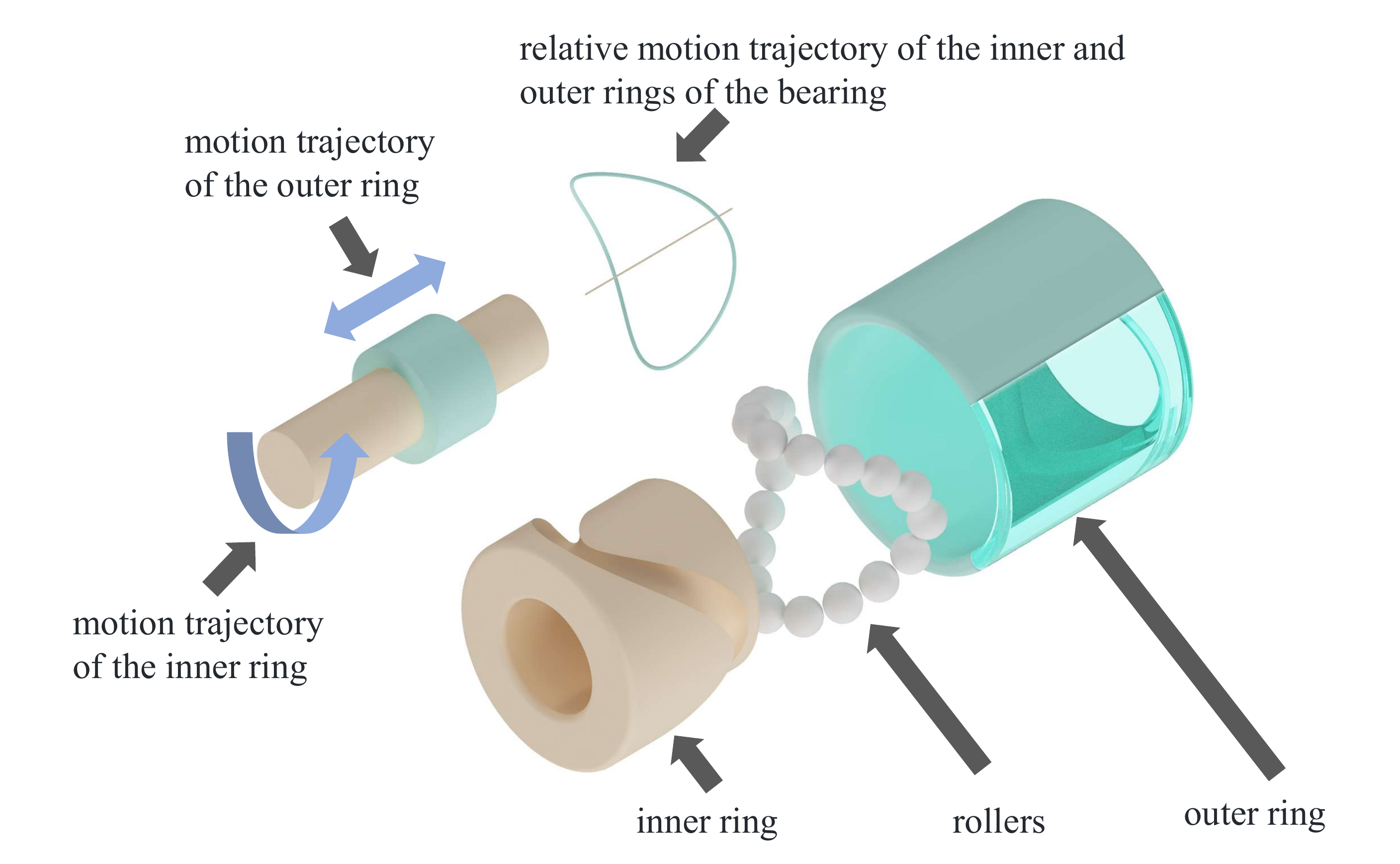}
	\caption{Structure and movement trajectories of components of CGBBR}
	\label{fig:bearing_move}
\end{figure}

The structure of CGBBR is shown in Fig. \ref{fig:bearing_move}, consisting of one outer ring, one inner ring, and one set of 
rolling elements. Unlike conventional ball bearings, the curved groove of this 
bearing's outer ring constitutes a spatial surface, generated by traversing a quarter-circle 
arc along a closed, smooth curve with a specified amplitude and period\cite{guoCurvedGrooveBall2021}. Likewise, the inner 
ring's curved groove represents a spatial surface created by progressing a three-quarter-circle 
arc in accordance with a closed, smooth curve, also exhibiting a defined amplitude and period.
Rolling elements are distributed in the curved groove of the inner ring. 
When the inner ring remains stationary and the curved groove on the outer ring interacts 
with the rolling elements at various positions, the outer ring must execute a reciprocating 
motion while rotating. When the outer ring does not rotate and the inner ring does not move axially, the outer ring will reciprocate upon the rotation of the inner ring. Similarly, the active and passive sequence of the inner and outer ring 
can be interchanged. This interchangeability provides more flexibility and opportunities for experimentation and innovation. 
% When the inner ring remains stationary and the curved groove on the outer ring interacts 
% with the rolling elements at various positions, the outer ring must execute a reciprocating 
% motion while rotating. If the rotational and axial movement of the outer ring and inner ring are 
% constrained, respectively, the outer ring exhibits a reciprocating motion as it revolves 
% around the inner ring. Similarly, the active and passive sequence of the inner and outer ring 
% can be interchanged. This results in one of the rings being able to act as an active element 
% or a passive element which interchangeability provides more flexibility and opportunities for 
% experimentation and innovation. 

CGBBR exhibits substantial differences from conventional bearings. In ordinary bearings, 
the raceway centerline projection onto the radial plane assumes a circular shape, 
whereas the raceway center curves for both inner ring and outer ring of 
CGBBR form closed, smooth three-dimensional curves, also projecting circularly onto the radial 
plane. Furthermore, the cage in CGBBR has been eliminated, and the rolling 
elements are completely mounted in the inner ring raceway. Consequently, traditional bearings' 
design methods, kinematic properties, and vibration characteristics become unsuitable for CGBBR. 
Therefore, it is imperative to analyze the structure, design principles, kinematic laws, and 
dynamic response characteristics of CGBBR. This comprehensive analysis will provide a 
theoretical basis for the optimization of the structure, standardization of production, and 
future widespread implementation of CGBBR.

There are three main contributions in this paper. Firstly, this paper proposes a novel type of 
bearing called CGBBR and the design method of it, which is named closed curve envelopment theory. 
The feasibility of the method is verified by manufacturing a real CGBBR, demonstrating 
that CGBBR can completely realize motion conversion. Secondly, the diameter-to-stroke ratio is 
proposed as the key parameter for optimal design of CGBBR. Finally,
vibration signal of CGBBR is analyzed using time-frequency analysis to identify the presence of 
periodic pulse clusters resulting from rolling element collisions. The analysis reveals that the 
collisions occur when the inner ring and outer ring raceway center curves align in the same direction.

The remainder of this paper is organized as follows: Section \ref{sec:Structural design of CGBBR} 
presents the design principles and structural features of CGBBR, along with a proposed design 
scheme; Section \ref{sec:Parameter analysis and kinematic analysis of CGBBR} examines the design 
parameters, the motion equations for any point on the bearing, and 
the rolling element's motion equation; Section \ref{sec:Kinematic verification of CGBBR} constructs a physical 
prototype of CGBBR, validating the design method's feasibility 
and the bearing function's comprehensiveness; Section \ref{sec:vibration analysis} investigates 
the actual vibration signals of CGBBR, analyzing its 
characteristics and underlying causes; Finally, conclusions are provided in Section \ref{sec:conclusion}.

\section{Structural design of CGBBR}
\label{sec:Structural design of CGBBR}

\subsection{Design background}
\label{subsec:sturctual design}

In transmission mechanisms, spatial curves are essential for converting rotary motion into 
reciprocating motion, as they constrain the trajectory of driven components, allowing for 
control and adjustment of their motion via the curve's geometry\cite{uickerTheoryMachinesMechanisms2017}. 
For example, spatial cam employs spatial curve to regulate the motion of driven components. While the 
cam acts as a driving component, it rotates to lead the driven component along the curved path, facilitating 
reciprocating motion\cite{chengExact3DPath2022,chengSpatialtemporalMotionControl2021}. 
Similarly, in ball screw mechanisms, balls roll along a spatial helical 
curve in the raceway\cite{huKinematicsBallscrewMechanisms2014}, propelling the nut into linear reciprocating motion. 
The mentioned mechanisms share a common feature: the axial projection of the driven component's motion along 
the spatial curve forms a reciprocating straight line. Simultaneously, mechanisms employed 
must control and stabilize the motion of the follower along the curved trajectory. CGBBR features
inner and outer races with curved raceways based on spatial curves, with rolling elements 
fully installed in the inner raceway, enabling the interchange between rotary and reciprocating 
motions. The aforementioned features of the motion conversion mechanisms are applied to the 
design of CGBBR. We optimized the structure by guiding the follower's movement along a 
sinusoidal curve.

\subsection{Design theory}
\label{subsec:Design theory}

CGBBR combines the structural features of space cam, ball screw and conventional bearings, 
making the traditional design method insufficient for meeting its structural requirements.
Design of CGBBR is developed utilizing closed curve envelopment theory, which combines the envelope theory
\cite{backhouseEnvelopeTheoryApplied1990,tsayApplicationTheoryEnvelope1994,changGeneralFrameworkGeometry2009,dhandeUnifiedApproachAnalytical1975} 
with swept surface theory\cite{gonzalez-palaciosGenerationContactSurfaces1994,hwangProfileSurfacesCylindrical2009}. 
This approach calculates the space envelope of a 3D curve trajectory. The surfaces of inner ring and outer ring of CGBBR are formed 
by sweeping a spherical surface based on both the 3D curve and its envelope.

% The design of the CGBBR is completed by integrating the envelope theory
 
% and swept surface theory of spatial cam 
% design. According to the envelope theory, cam surface is defined as an envelope for the follower's surface at different positions during a complete cycle of cam rotation.
% Based on the sweep surface theory, the groove surface of cam is obtained by computing the swept 
% surface of the follower in motion.

\begin{figure}[htbp]
	\centering
	\includegraphics[width=1\textwidth]{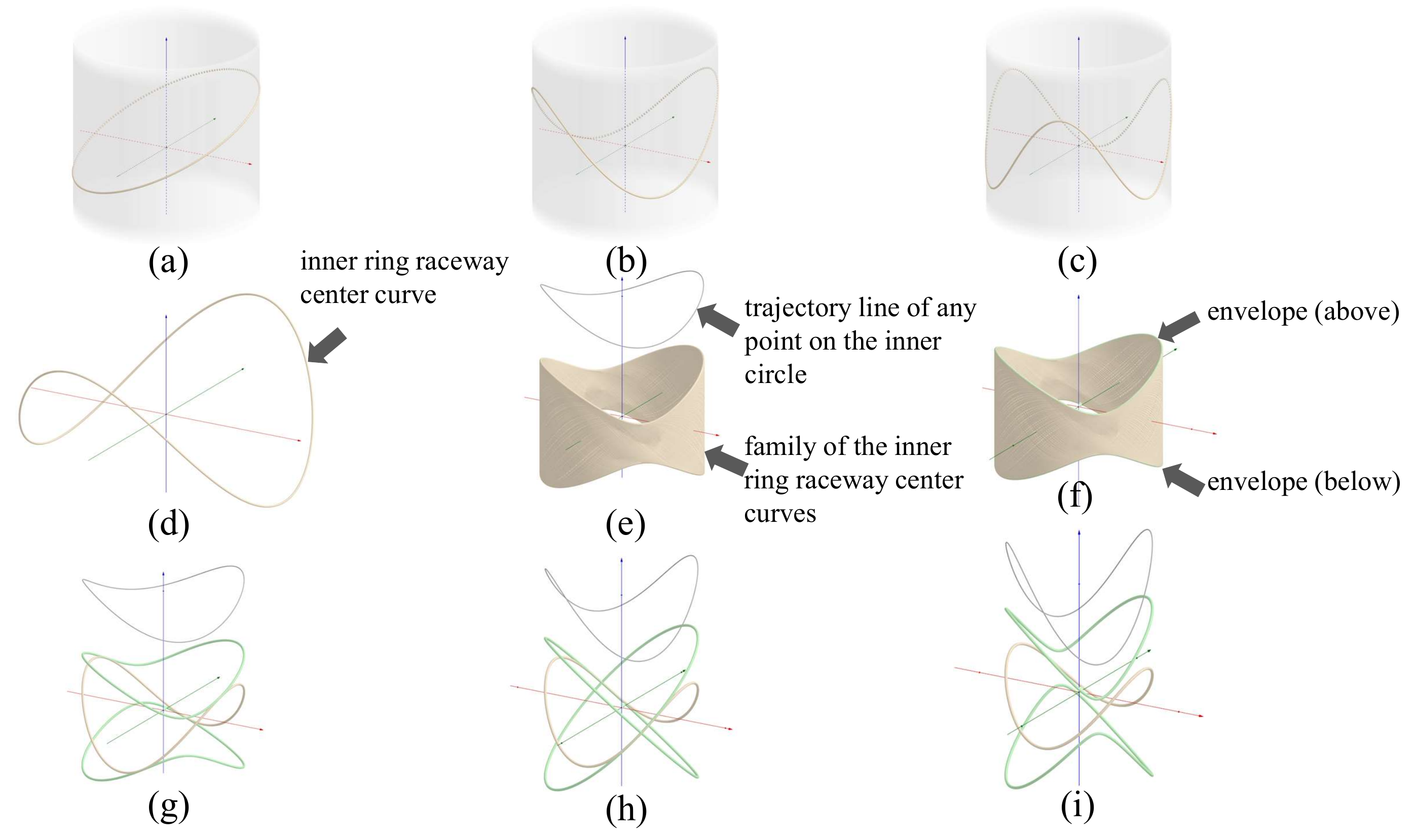}
	\caption{ Generation process of IRRCC and ORRCC. (a) IRRCC $n_1=1$; (b) IRRCC $n_1=2$; 
    (c) IRRCC $n_1$=3; (d) inner ring raceway center curve $n_1=2$; (e)  motion trajectory and 
    family of IRRCCs; (f) envelopes of IRRCC; (g)ORRCC $0<m<1$; (h) ORRCC $m=1$; (i) ORRCC $m>1$.}
	\label{fig:curve_and_envelope}
\end{figure}

Firstly, the function of envelope theory in CGBBR design is expounded. The inner ring raceway 
center curve (IRRCC) of the CGBBR is shifted one circle along the aforementioned sinusoidal 
displacement curve to produce a family of spatial curves. The envelope of this spatial curve 
family is the outer ring raceway center curve (ORRCC). IRRCC is a closed, smooth, and 
non-self-intersecting 3D curve that surrounds the bearing's rotation axis\cite{chengSpatialtemporalMotionControl2021}, 
which is defined by the Equation (\ref{eq:sin}):
\begin{equation}
    \left\{ \begin{array}{l}
        x\,\,=\,\,R\cos t\\
        y\,\,=\,\,R\sin t\\
        z\,\,=\,\,A\cos n_1t\\
    \end{array} \right. 
    \label{eq:sin}
\end{equation}
IRRCC is a wavy line on a cylindrical surface, which, when divided along its ``peaks'' or 
``valleys'', can be unfolded into a cosine curve on a plane\cite{guoCurvedGrooveBall2021}. 
In Equation (\ref{eq:sin}), $A$ represents the amplitude of the cosine curve. $R$ represents the 
radius of the cylindrical surface enclosed by IRRCC. $n_1$ is the number of ``peaks'' or ``valleys'' 
of IRRCC. Subfigures (a)-(c)of Fig. \ref{fig:curve_and_envelope} indicate the above relationships.

As the displacement curve of the follower is sinusoidal, the trajectory of any point on the 
inner ring with respect to the outer ring can be expressed as Equation (\ref{eq:move_trajectory}):
\begin{equation}
    \left\{ \begin{array}{l}
        x=R\cos t\\
        y=R\sin t\\
        z=mA\sin n_2t\\
    \end{array} \right. 
    \label{eq:move_trajectory}
\end{equation}

where $mA$ is the motion amplitude of IRRCC, $n_2$ is the number of IRRCC reciprocal movements 
in one rotation time. Since the number of ``peaks'' and ``valleys'' of the inner ring and the 
outer ring of CGBBR is equal, $n_1=n_2$.

The space curve family formed by the IRRCC, rotating along its trajectory, can be represented as a single-parameter space curve family denoted by the equation $f(x,y,z,C)=0$, which can be specifically written as:
% The family of spatial curves that are formed by rotating IRRCC along the trajectory can be 
% expressed by the equation $f\left( x,y,z,C \right) =0$, which can be specifically written as:
\begin{equation}
    \left\{ \begin{array}{l}
        x=R\cos t\\
        y=R\sin t\\
        z=A\cos\mathrm{(}n(t-C))+mA\sin(nC)\\
    \end{array} \right. 
    \label{eq:family of curve}
\end{equation}
where $n=n_1=n_2$, $C$ is a parameter. The family of spatial curves is shown in subfigure 
(e) of Fig. \ref{fig:curve_and_envelope}.

Envelopes of the family of the spatial curves are ORRCC, which can be expressed as:
\begin{equation}
    \left\{ \begin{array}{c}
        f\left( x,y,z,C \right) =0\\
        f'_C\left( x,y,z,C \right) =0\\
    \end{array} \right. 
    \label{eq:System of equations}
\end{equation}
Due to the presence of ``peaks'' and ``valleys'' in IRRCC, distinct envelope lines are generated 
when moving along a sine trajectory. The ``peaks'' and ``valleys'' each produce their own envelope 
lines, which can be expressed in Equation (\ref{eq:envelope1}) and Equation (\ref{eq:envelope2}) respectively.
\begin{equation}
    \left\{ \begin{array}{c}
        x=R\cos t\\
        y=R\sin t\\
        z=A\cos \left[ n\left( t+\frac{\mathrm{arc}\tan \left( \frac{m\tan ^2\left( n\frac{t}{2} \right) +m+2\tan \left( n\frac{t}{2} \right)}{\tan ^2\left( n\frac{t}{2} \right) -1} \right)}{n} \right) \right]\\
        -mA\sin \left[ \mathrm{arc}\tan \left( \frac{m\tan ^2\left( n\frac{t}{2} \right) +m+2\tan \left( n\frac{t}{2} \right)}{\tan ^2\left( n\frac{t}{2} \right) -1} \right) \right]\\
    \end{array} \right. 
    \label{eq:envelope1}
\end{equation}
\begin{equation}
    \left\{ \begin{array}{c}
        x=R\cos t\\
        y=R\sin t\\
        z=A\cos \left[ n\left( t-\frac{\pi -\mathrm{arc}\tan \left( \frac{m\tan ^2\left( n\frac{t}{2} \right) +m+2\tan \left( n\frac{t}{2} \right)}{\tan ^2\left( n\frac{t}{2} \right) -1} \right)}{n} \right) \right]\\
        +mA\sin \left[ \mathrm{arc}\tan \left( \frac{m\tan ^2\left( n\frac{t}{2} \right) +m+2\tan \left( n\frac{t}{2} \right)}{\tan ^2\left( n\frac{t}{2} \right) -1} \right) \right]\\
    \end{array} \right. 
    \label{eq:envelope2}
\end{equation}
Subfigures (d)-(f) in Fig. \ref{fig:curve_and_envelope} demonstrate the application of 
envelope theory in the design of CGBBR.

$m$ in Equations (\ref{eq:move_trajectory})-(\ref{eq:envelope2}) determines
the contact conditions of upper ORRCC and lower ORRCC.

As shown in subfigures (g)-(i) of Fig. \ref{fig:curve_and_envelope}: When $0<m<1$, IRRCC is 
tangent to upper ORRCC and lower ORRCC. When $m=1$, IRRCC is tangent to upper ORRCC and lower ORRCC. Upper ORRCC intersects with lower ORRCC. When $m>1$, IRRCC can be tangent to either the 
upper or lower ORRCC, but not both simultaneously.

Based on Equations (\ref{eq:sin})-(\ref{eq:envelope2}) and Fig. \ref{fig:curve_and_envelope},
we define ``single reciprocating movement'' as the motion in which the inner ring moves away from the initial position, and then returns to the initial position. The following relationship can be obtained:
\begin{itemize}
    \item A single rotation of CGBBR corresponds to $n$ reciprocating motions.
    \item A reciprocating motion corresponds to an angle change of $\frac{2\pi}{n}$
    \item The maximum displacement of a single reciprocating motion is $2mA$
\end{itemize}

\begin{figure}[htbp]
	\centering
	\includegraphics[width=0.8\textwidth]{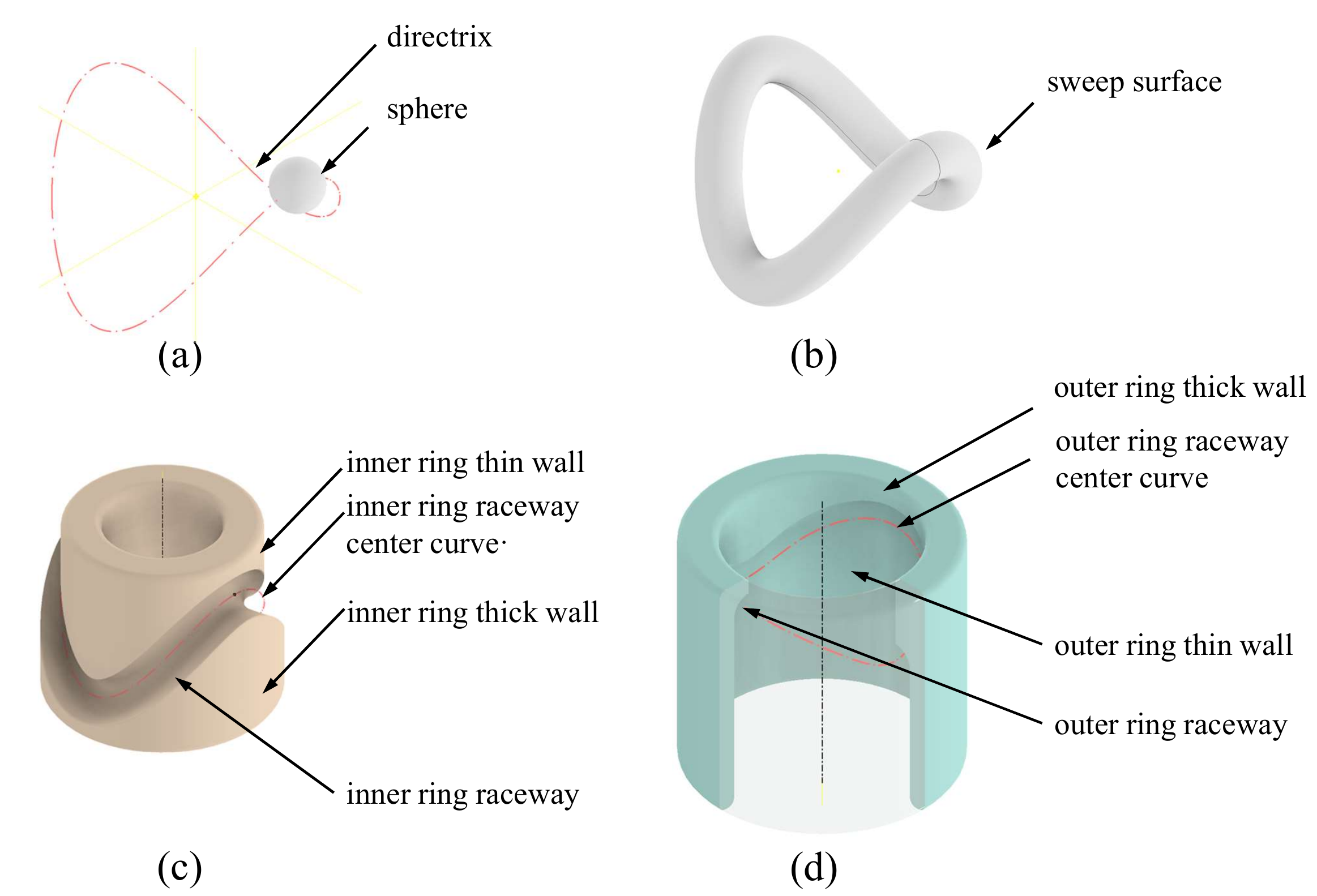}
	\caption{Design principles and structures of the inner ring and outer ring. (a)elements of the 
    swept surface; (b)swept surface; (c)structure of inner ring; (d)structure of outer ring.}
	\label{fig:sweep_surface}
\end{figure}

The curved surface raceways, accomplished through the intricate swept surface theory, 
constitute the fundamental architecture of both the inner ring and outer ring of CGBBR.
For instance, as illustrated in Fig. \ref{fig:sweep_surface} (a), during the design 
process of the inner race groove surface, the rolling element moves along the IRRCC for one revolution, resulting in the development of the curved surface 
of the inner race groove which is called the swept surface. The swept surface is depicted in 
subfigure (b) of Fig. \ref{fig:sweep_surface}. Equation (\ref{eq:sweep_surface}) can be used to 
represent the swept surface.
\begin{equation}
    \left\{ \begin{array}{l}
	    x=x\left( t \right) +r\cos \phi \sin \theta\\
	    y=y\left( t \right) +r\sin \phi \sin \theta\\
	    z=z\left( t \right) +r\cos \theta\\
    \end{array} \right. 
    \label{eq:sweep_surface}
\end{equation}
where $r$ is the radius of rolling element. $\theta$ and $\phi$ are the polar angle and azimuth angle in the spherical coordinate 
system, which can be calculated using the following formulas.
\begin{equation}
\theta =\mathrm{arc}\tan \left( \frac{\sqrt{x'^2+y'^2}}{z'^2} \right) , \phi =\mathrm{arc}\tan \left( \frac{y'}{x'} \right) 
  \label{theta}
\end{equation}
where $x' = \frac{dx}{dt}$, $y' = \frac{dy}{dx}$, and $z' = \frac{dz}{dt}$ respectively 
represent the derivatives in the three directions at time $t$.

\begin{figure}[htbp]
	\centering
	\includegraphics[width=0.8\textwidth]{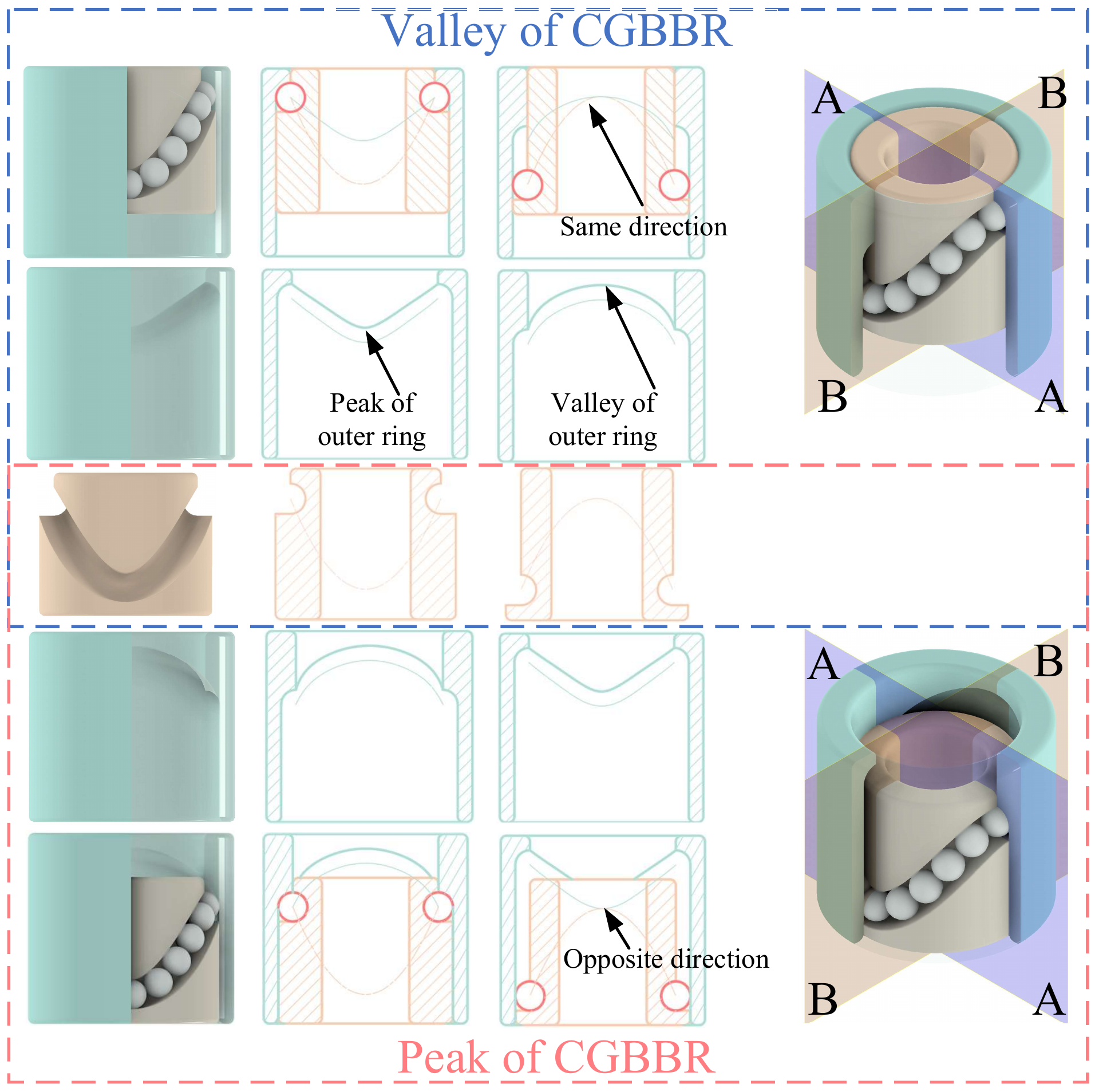}
	\caption{Columns from left to right represent: A-A Front View, 
    A-A Sectional View, B-B Sectional View, ``valley'' and ``peak'' of CGBBR.}
	\label{fig:v and p}
\end{figure}

Subfigures (c) and (d) in Fig. \ref{fig:sweep_surface} depict the CGBBR structure designed 
using the envelope theory and swept surface theory. Primary components of the inner ring 
consist of the inner ring's thin wall, thick wall, raceway, and the centerline of raceway.
Radius of inner ring thin wall is equal to the radius $R$ of IRRCC. Inner ring 
thin wall is separated from inner ring thick wall by inner ring raceway. Inner ring raceway 
is formed by subtracting the swept surface from inner ring using boolean operations.
Similarly, main structures of outer ring include: outer ring thin wall, outer ring thick wall, 
outer ring raceway, and outer ring raceway centerline. Radius of the outer ring thick wall is 
equal to the radius $R$ of ORRCC. The outer ring raceway separates the outer ring thick wall 
from the outer ring thin wall. Outer ring raceway is constructed through a boolean 
subtraction of the swept surface from the outer ring, thus creating a precisely engineered 
profile. Given that the radius of IRRCC matches that of ORRCC, radius of inner ring 
thin wall and outer ring thick wall are identical, and so are those of inner ring thick wall 
and outer ring thin wall.

Apart from the distinct component structure, CGBBR also possesses a significantly different 
assembly structure compared to ordinary bearings, which is mainly due to the structure with 
curvatures in its inner ring and outer ring. We will demonstrate this using the operational 
process of a CGBBR that has two ``peaks'' and two ``valleys''.
Fig. \ref{fig:v and p} shows a $\frac{1}{4}$ transparency of the outer ring of CGBBR to 
enable a clear observation of the contact and positional relationships of each component during 
its movement. The operation of CGBBR involves constant positional changes of the inner and outer rings.
To differentiate between these positions, the following nomenclature will be utilized:
the sharp points on the ORRCC correspond to the ``peaks'' situated on the outer ring, whereas 
relatively smooth points correspond to the ``valleys'' located on the outer ring. 
The ``valleys'' in CGBBR is formed by the alignment of IRRCC and ORRCC in the same direction, while the 
``peaks'' is formed when they are aligned in opposite directions. The alternation of ``peaks'' and 
``valleys'' in CGBBR affects the angular and positional relationships between its inner ring 
and outer ring, which explains its capability to transition between rotary and reciprocating 
motion.

\section{Parameter analysis and kinematic analysis of CGBBR}
\label{sec:Parameter analysis and kinematic analysis of CGBBR}

\subsection{Diameter stroke ratio}
\label{subsec:Diameter stroke ratio}

The motion conversion ability of CGBBR is directly influenced by the relative motion 
trajectory equation of the bearing's  inner ring and outer ring. As defined in Equation (\ref{eq:move_trajectory}),
since the curve transforms into a cosine curve after being unfolded, its steepness and 
smoothness are directly affected by the period and amplitude. Accordingly, it is imperative to make appropriate adjustments to these two variables. 
Equation (\ref{eq:slider_crank}) displays the displacement 
curve of the engine's crank connecting rod mechanism, which can be simplified into a cosine curve, as shown in Equation (\ref{eq:simplify_slider_crank}).
\begin{equation}
    x=r\cos \alpha +\sqrt{l^2-\sin ^2\alpha}
    \label{eq:slider_crank}
\end{equation}
\begin{equation}
    x=r\cos \alpha +l
    \label{eq:simplify_slider_crank}
\end{equation}
\begin{figure}[htbp]
	\centering
	\includegraphics[width=1\textwidth]{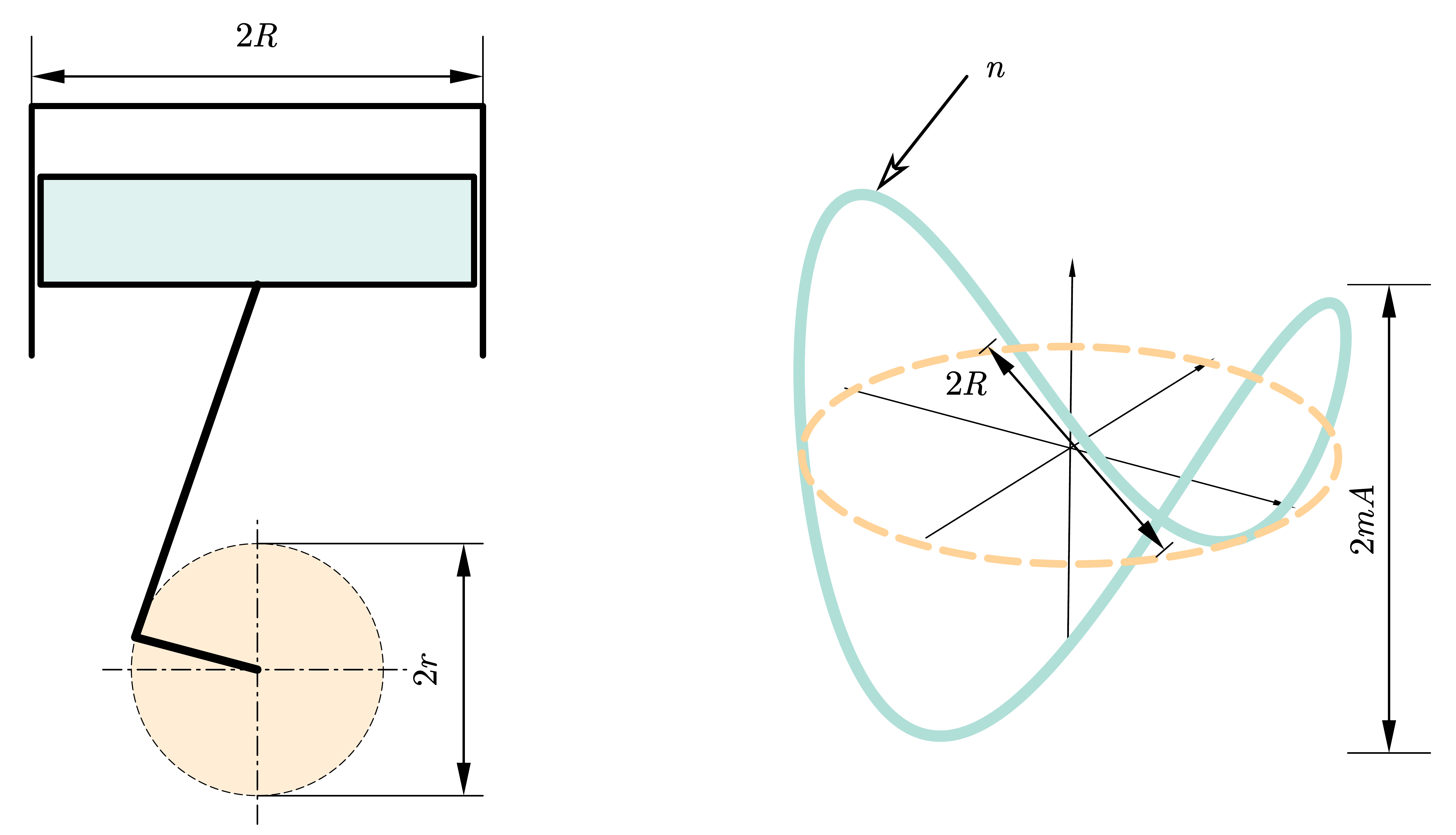}
	\caption{Motion diagram of crank connecting rod and CGBBR}
	\label{fig:slider_crank_and_curve}
\end{figure}
where $x$ is the distance of the end of the connecting rod from the crank axle, $l$ is the 
length of the connecting rod, $r$ is the length of the crank, and $\alpha$ is the angle of 
the crank measured from top dead center\cite{uickerTheoryMachinesMechanisms2017}.
In balancing the effect of the relationship between period and amplitude in the displacement 
curve, the crank-connecting rod mechanism employs an appropriate cylinder bore to stroke ratio($B/S$).
The $B/S$ of engines currently ranges from 0.887 to 1.342\cite{ozcanPerformanceEmissionCharacteristics2008,yaminPerformanceSimulationFourstroke2004a}.
When the crank connecting rod mechanism completes one rotation, the piston reciprocates once. 
Similarly, as the CGBBR makes one revolution, the inner and outer rings move $n$ times relative 
to each other. With reference to $B/S$, we proposed the definition of  diameter stroke 
ratio($D/S$) of CGBBR, which represents the ratio between diameter of the trajectory curve and the stroke.
By utilizing the parameters depicted in Fig. \ref{fig:slider_crank_and_curve}, 
\begin{figure}[htbp]
	\centering
	\includegraphics[width=1\textwidth]{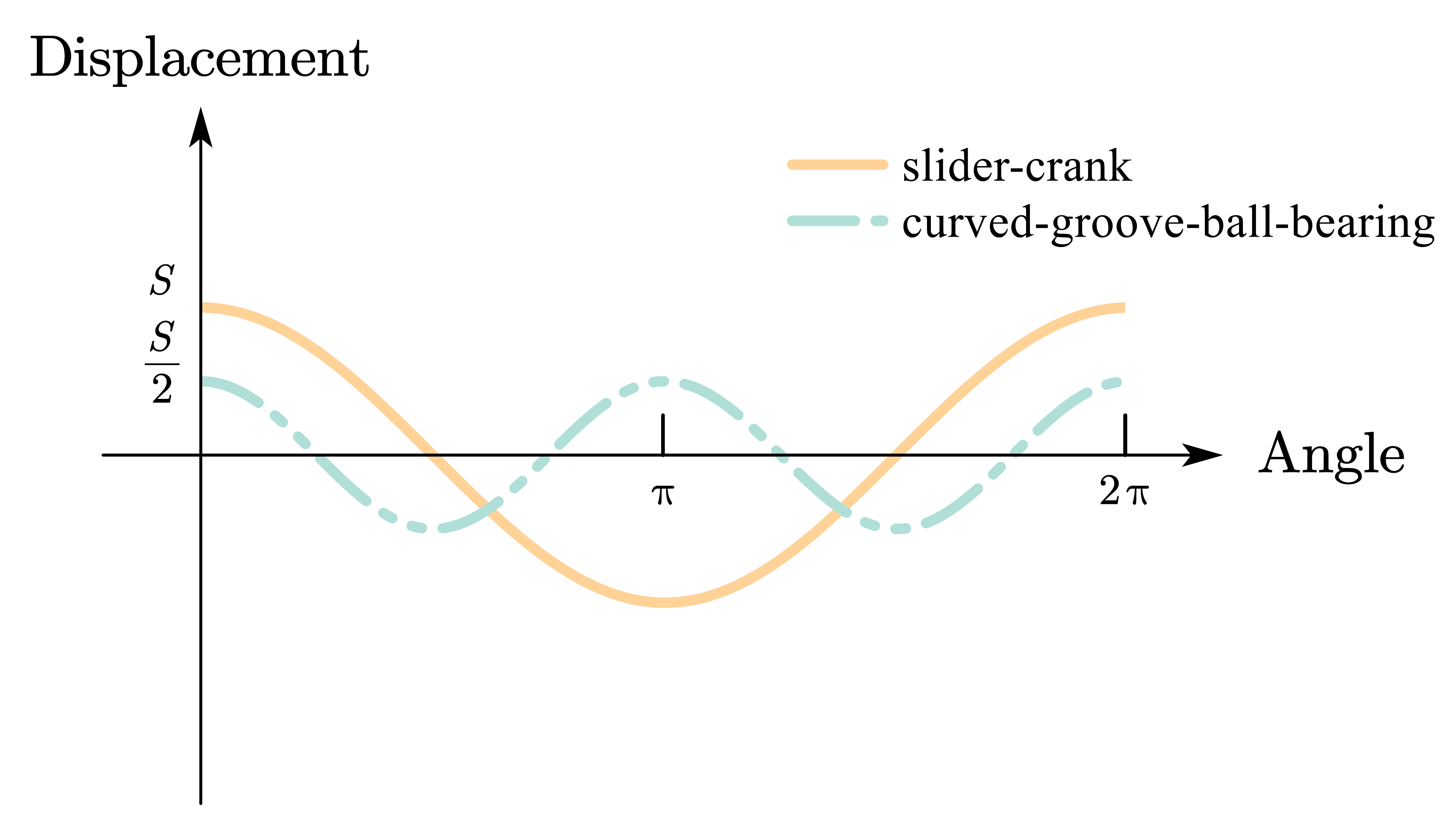}
	\caption{Comparison of displacement of crank connecting rod and CGBBR with angle}
	\label{fig:curve_scaling}
\end{figure}
$B/S=\frac{2R}{2r}$ and $D/S=\frac{2R}{2mnA}$. Alterations to $n$, the number of reciprocating 
motions for each revolution of the CGBBR, will have an impact on its stroke. 
To ensure the stability of $D/S$, modifications to $mA$ are required. This entails altering at 
least one parameter either in $m$ or $A$. The stroke of the crank-connecting rod mechanism is 
$n$ times the stroke of CGBBR when $B/S$ and $D/S$ are identical. The displacement of the crank-connecting 
rod and CGBBR with respect to the rotational angle is compared in Fig. \ref{fig:curve_scaling},
where $B/S$ is equal to $D/S$ and $n$ is equal to 2.

\subsection{Calculation of the number of rolling elements of CGBBR}
\label{ball_optimized}

\begin{figure}[htbp]
	\centering
	\includegraphics[width=1\textwidth]{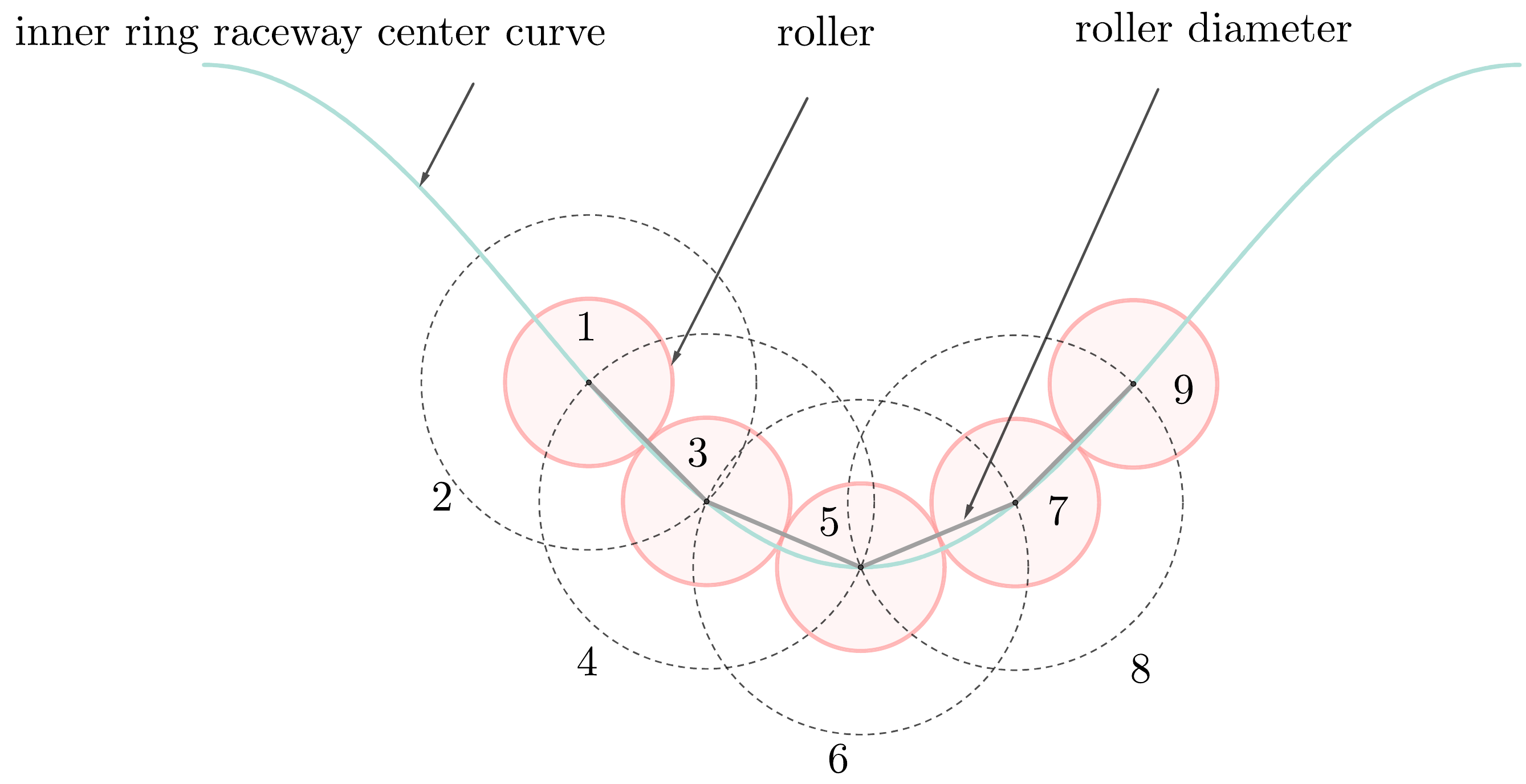}
	\caption{Calculation process of the number of rolling elements. The red solid circles in the  
    diagram represent rolling elements, and the dashed circles represent spherical surfaces. 
    Number labels indicate the order of calculation, and the numbers of odd number calculated 
    represents the number of rolling bodies.}
	\label{fig:number_of_the_roller}
\end{figure} 

The rolling elements of CGBBR are fully filled in the curved groove raceway of the inner 
ring, which is different from the structure of ordinary bearings.
As such, it is necessary to determine the precise number of rolling elements for the CGBBR. 
This paper presents a methodology for accomplishing this task.
Assuming that the rolling elements are rigid and that adjacent rolling elements are in 
tangent with each other, the centers of these rolling elements are positioned on IRRCC.
Calculating the number of rolling elements involves the following steps:
\begin{itemize}
    \item Select a point on IRRCC as the center of the first rolling element and form a spherical surface with the rolling element diameter as the radius. By solving the equations of the spherical surface and the center curve of the raceway, determine two intersection points.
    \item Use one of the intersection points as the center of the second rolling element and form another spherical surface with the rolling element diameter as the radius. By solving the equations of the second spherical surface and IRRCC, determine two intersection points. Select the intersection point on the vector connecting the centers of the first and second rolling elements as the position of the center of the next rolling element.
    \item Repeat the above steps, using the current center of the rolling element as the starting point, forming a new spherical surface, and solving the equations to determine the intersection point, until the distance between the intersection point and the line connecting the center of the first rolling element is less than the radius of the rolling element.
    \item Calculate the total number of rolling elements.
\end{itemize}
To facilitate understanding, we utilize the form of a schematic diagram to describe the process 
described above, as shown in Fig. \ref{fig:number_of_the_roller}.

\subsection{Kinematic analysis of CGBBR}
\label{kinematic analysis}

CGBBR has the capability to switch between rotational and reciprocating motion. Regardless of 
whether the inner ring or outer ring acts as the driving component, the fundamental characteristic 
of its motion is that the cylinder undergoes both rotational and reciprocating motion 
along its own axis simultaneously. Therefore, the motion model of CGBBR can be conceptualized as the trajectory of any point on a cylinder during its simultaneous rotation and reciprocating motion. The motion can be decomposed into two independent components: rotational and reciprocating. 
These components can be described using homogeneous transformation matrices\cite{huKinematicsBallscrewMechanisms2014}.
Rotational motion is performed around the axis of the bearing, represented by the rotational 
matrix $R(t)$ assuming an angular velocity of $\omega$. The equation for periodic motion can be 
expressed as $z=2mA\sin nt$ according to Equation (\ref{eq:move_trajectory}). This equation 
can also be represented by the periodic matrix $T(t)$.
Compound motion is the combination of rotational and reciprocating motions, which can be 
represented as a matrix multiplication of rotation matrix and translation matrix. The matrix 
representation of reciprocating and rotational motion is: 
\begin{equation}
    H\left( t \right) =T\left( t \right) \cdot R\left( t \right) =\left[ \begin{matrix}
        \cos \omega t&		-\sin \omega t&		0&		0\\
        \sin \omega t&		\cos \omega t&		0&		0\\
        0&		0&		1&		A\sin nt\\
        0&		0&		0&		1\\
    \end{matrix} \right] 
    \label{eq:HT}
\end{equation}
The motion of any point on the inner ring or outer ring can be represented using the matrix 
$H(t)$.
Assuming that the rolling element can be approximated as a mass point moving on the inner 
ring raceway, its trajectory can be represented as movement along the IRRCC.
Represented by a matrix:
\begin{equation}
    p_{local}\left( t \right) =\left[ \begin{array}{c}
        R\cos t\\
        R\sin t \\
        A\cos nt \\
    \end{array} \right] 
    \label{eq:plocal}
\end{equation}
The homogeneous transformation matrix $H(t)$ representing the motion of the cylinder can be 
utilized to calculate the rolling element's motion equation by multiplying with the mass 
point coordinates of the local coordinate system. To obtain the displacement equation of the 
rolling element, the Equation (\ref{eq:move_trajectory}) must first be expanded into a 
homogeneous coordinate form. Then, the resulting equation is multiplied by the homogeneous 
transformation matrix $H\left( t \right)$.
\begin{equation}
    p\left( t \right) =\left[ \begin{array}{c}
        R\cos \left( \omega t \right) \cos \left( t \right) -R\sin \left( \omega t \right) \sin \left( t \right)\\
        R\sin \left( \omega t \right) \cos \left( t \right) +R\cos \left( \omega t \right) \sin \left( t \right)\\
        A\cos \left( nt \right) +A\sin \left( nt \right)\\
        1\\
    \end{array} \right] 
    \label{eq:pt}
\end{equation}
Converting this to Cartesian coordinates is:
\begin{equation}
    p\left( t \right) =\left[ \begin{array}{c}
        R\cos \left( \omega t \right) \cos \left( t \right) -R\sin \left( \omega t \right) \sin \left( t \right)\\
        R\sin \left( \omega t \right) \cos \left( t \right) +R\cos \left( \omega t \right) \sin \left( t \right)\\
        A\cos \left( nt \right) +A\sin \left( nt \right)\\
    \end{array} \right] 
    \label{eq:PT}
\end{equation}

\section{Kinematic verification of CGBBR}
\label{sec:Kinematic verification of CGBBR}

Design method of CGBBR proposed in Section \ref{sec:Structural design of CGBBR} will be experimentally 
validated in this section, alongside functionalities of CGBBR. Specifically, this section aims 
to verify the capability of CGBBR to convert rotary motion to reciprocating motion and vice versa, 
as well as the function of compound motion. The CGBBR parameters used in the experiment are 
shown in the Table \ref{tab:Parameters of CGBBR}:

% \begin{table}[htbp]
%     \centering
%     \caption{Parameters of CGBBR}
%     \label{tab:Parameters of CGBBR}
%     \begin{tabular}{cc}
%       \toprule
%       Parameter/unit & Value \\
%       \midrule
%       Raceway Center Curve Radius $R$/mm & 25 \\
%       Raceway Center Curve Amplitude $A$/mm & 25 \\
%       Reciprocating amplitude $mA$/mm & 12.5 \\
%       Reciprocating times per revolution $n$/times & 2 \\
%       Roller element diameter/mm & 10 \\
%       Number of rolling elements/number & 20 \\
%       \bottomrule
%     \end{tabular}
% \end{table}

\begin{table}[htbp]
    \centering
    \caption{Parameters of CGBBR}
    \label{tab:Parameters of CGBBR}
    \begin{tabular}{l l}
      \toprule
      Parameter/unit & Value \\
      \midrule
      Raceway Center Curve Radius $R$/mm & 25 \\
      Raceway Center Curve Amplitude $A$/mm & 25 \\
      Reciprocating amplitude $mA$/mm & 12.5 \\
      Reciprocating times per revolution $n$/times & 2 \\
      Roller element diameter/mm & 10 \\
      Number of rolling elements/number & 20 \\
      \bottomrule
    \end{tabular}
\end{table}

\begin{figure}[htbp]
	\centering
	\includegraphics[width=1\textwidth]{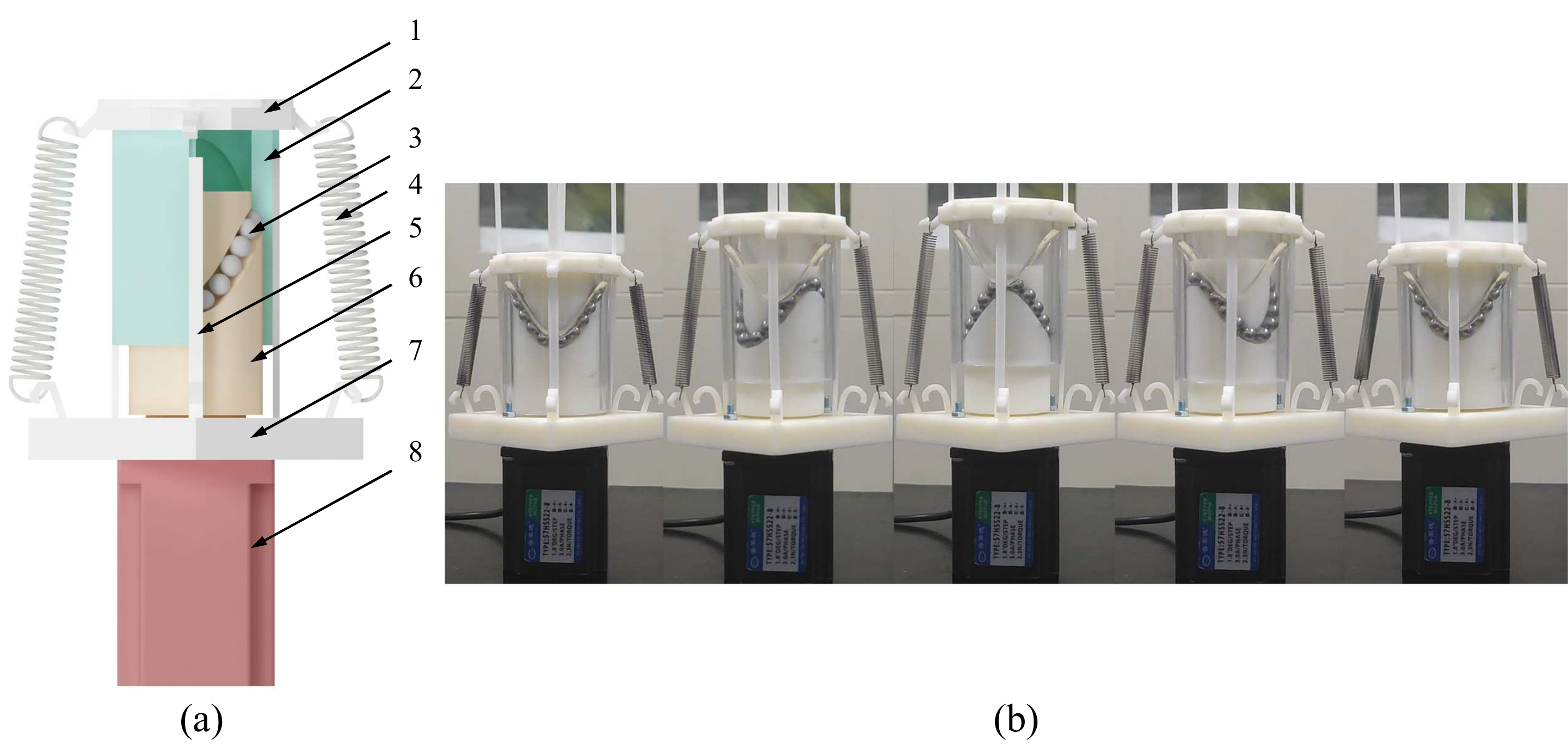}
	\caption{Verification of the ability of CGBBR to convert rotary motion to reciprocating motion.
    (a)composition of the experimental device: 1.lid; 2.outer ring; 3.rolling element; 4.spring; 
    5.limit bar; 6.inner ring; 7.pedestal; 8.stepper motor. (b)realization process from rotary motion to reciprocating motion:
    From left to right, ``valley'' of CGBBR, switching process of ``valley'' to ``peak'' of CGBBR, ``peak'' of CGBBR,
    switching process of ``peak'' to ``valley'' of CGBBR, ``valley'' of CGBBR.}
	\label{fig:experiment1}
\end{figure}

To assess the capacity of CGBBR in transforming rotary motion to reciprocating motion, we 
employed an actual object. Specifically, 3D printing technology was utilized to create the 
inner ring and transparent outer ring using photosensitive resin material\cite{chengSpatialtemporalMotionControl2021}. As depicted in 
Fig. \ref{fig:experiment1}.

We evaluated the performance of CGBBR by converting rotary motion of inner ring  to 
reciprocating motion of outer ring. We constrain the degrees of freedom of the inner ring and 
outer ring, allowing only rotary motion for the inner ring, while the outer 
ring is restricted to reciprocate. The implementation process entails positioning the initial 
position at the ``valley'' of the bearing, driving the inner ring rotation with the 
stepping motor, and applying pressure on the upper part of the outer ring using the spring.
The inner ring rotates as the outer ring moves upwards, continuously approaching bearing's ``peak''.
After reaching the ``peak'', the inner ring continues to rotate while the spring, raceway, 
and rolling elements work together to move the outer ring downward until it reaches the ``valley'' of
bearing. The transformation from rotational motion to reciprocating motion can be achieved repeatedly.
Please watch the accompanying video for demos.

\begin{figure}[htbp]
	\centering
	\includegraphics[width=1\textwidth]{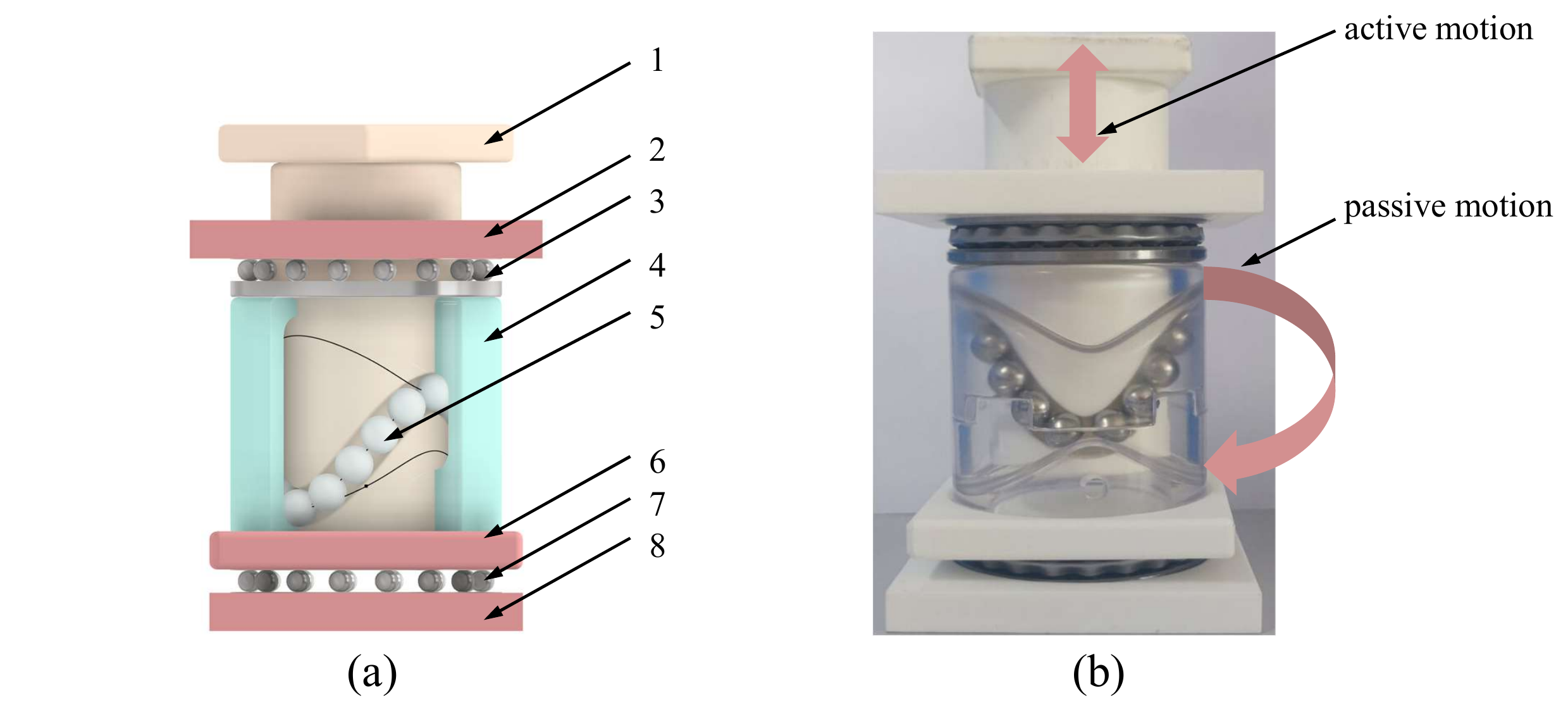}
	\caption{Verification of the ability of CGBBR to convert rotary motion to reciprocating motion. 
    (a)composition of the experimental device: 1.inner ring; 2.upper platen; 3.upper thrust ball bearing; 
    4.outer ring; 5.rolling element; 6.inertia board; 7.lower thrust ball bearing 8.lower platen; 
    (b)device movement diagram.}
	\label{fig:experiment2}
\end{figure}

The performance of CGBBR was assessed by transforming inner ring's reciprocating motion 
into outer ring's gyration\cite{guoCurvedGrooveBall2022}. As shown in Fig. \ref{fig:experiment2}, outer ring of  
CGBBR includes thrust ball bearings at its upper and lower ends. Therefore, outer ring is 
restricted to rotary motion. Applying a reciprocating force on the end 
face of inner ring will result in reciprocation of it, and subsequently push the 
outer ring to rotate. Please watch the accompanying video for demos.

\begin{figure}[htbp]
	\centering
	\includegraphics[width=1\textwidth]{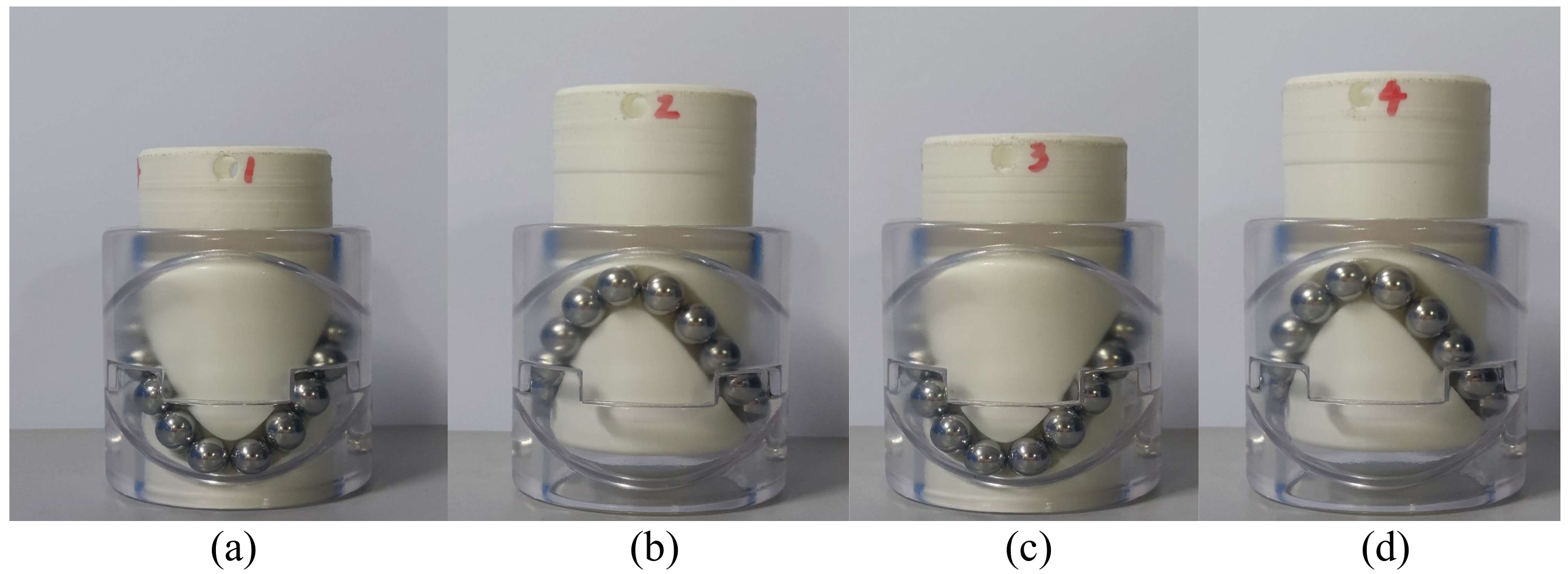}
	\caption{Verification of the ability of CGBBR to perform compound motion. (a)``valley'' of 
    CGBBR corresponds to number 1; (b)``peak'' of CGBBR corresponds to number 2; (c)``valley'' of 
    CGBBR corresponds to number 3; (d)``peak'' of CGBBR corresponds to number4.}
	\label{fig:experiment3}
\end{figure}

CGBBR's ability to achieve compound motion is based on the structure described in the 
literature\cite{guoCurvedGrooveBall2022}. The positions that are 90 degrees apart from the outer wall of the inner ring 
are marked with numbers 1-4, respectively. When a turning moment is applied to the inner ring 
while outer ring fixed, it will reciprocate along the spatial curve.
From a fixed perspective, it will be noticed that the height of the end surface of the 
inner ring varies in accordance with the alteration of the numbers inscribed on the outer 
surface of inner ring. please refer to the relevant information in Fig. \ref{fig:experiment3}.

The aforementioned three experiments demonstrate the feasible design of CGBBR, which 
combines the envelope theory and swept surface theory. They also indicate the full range of 
motion conversion capability of CGBBR.

\section{Vibration analysis of CGBBR}
\label{sec:vibration analysis}

\subsection{Experiment equipment}
\label{experimental equipment}

Due to an innovative curved raceway that contains rolling 
elements, CGBBR exhibits distinct vibrational properties compared to conventional 
bearings\cite{zhaoKinematicCharacteristicsFault2021,guoFaultCharacteristicFrequency2020}. To enhance our understanding and optimize the performance of CGBBR, it is essential 
to sample and extract vibration signals of this bearing.

\begin{figure}[htbp]
	\centering
	\includegraphics[width=1\textwidth]{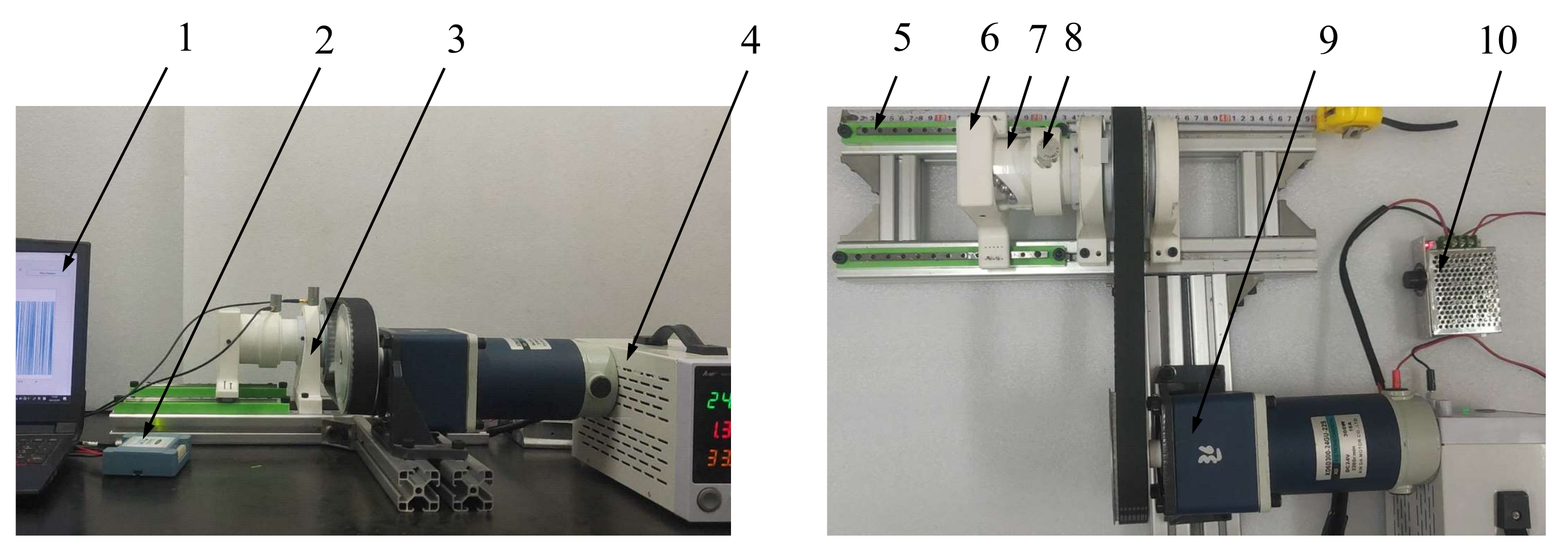}
	\caption{Signal acquisition platform of CGBBR. 1.computer; 2.signal acquisition card; 
    3.bearing box; 4.power supply; 5.guide rail; 6.sliding support mechanism; 7.CGBBR; 
    8.vibration sensor; 9.motor; 10.speed governor.}
	\label{fig:experimental_equipment}
\end{figure}

We constructed the acquisition platform for the vibration signals of CGBBR, as illustrated 
in Fig. \ref{fig:experimental_equipment}. The experimental platform functions based on the 
following working principle: the motor drives the driven shaft, which is rigidly coupled to 
inner ring. A sliding support mechanism is utilized to establish a connection between 
outer ring and the slide block of the guide rail. During the rotation of inner ring, outer 
ring undergoes a lateral displacement along the guide rail, while a vibration sensor affixed 
to the outer wall of outer ring concurrently captures vertical vibration signals emanating 
from the bearing. A signal acquisition system consisting of an MCC signal acquisition card 
and a computer is employed to capture and display the vibration signal emanating from CGBBR 
in real time.

The parameters of the CGBBR are listed in Table \ref{tab:Parameters of CGBBR}.
The IEPE piezoelectric accelerometer, CT1010LC, is utilized for measuring the vibration signal.
The data acquisition system of MCC is used to collect the vibration acceleration signal of 
CGBBR. To guarantee sampling accuracy, a sampling frequency of 25000 Hz and a sampling 
time of 20 seconds were selected.

\subsection{Vibration signal characteristics of CGBBR}
\label{sec:vibration signal analysis}

\begin{figure}[htbp]
	\centering
	\includegraphics[width=1\textwidth]{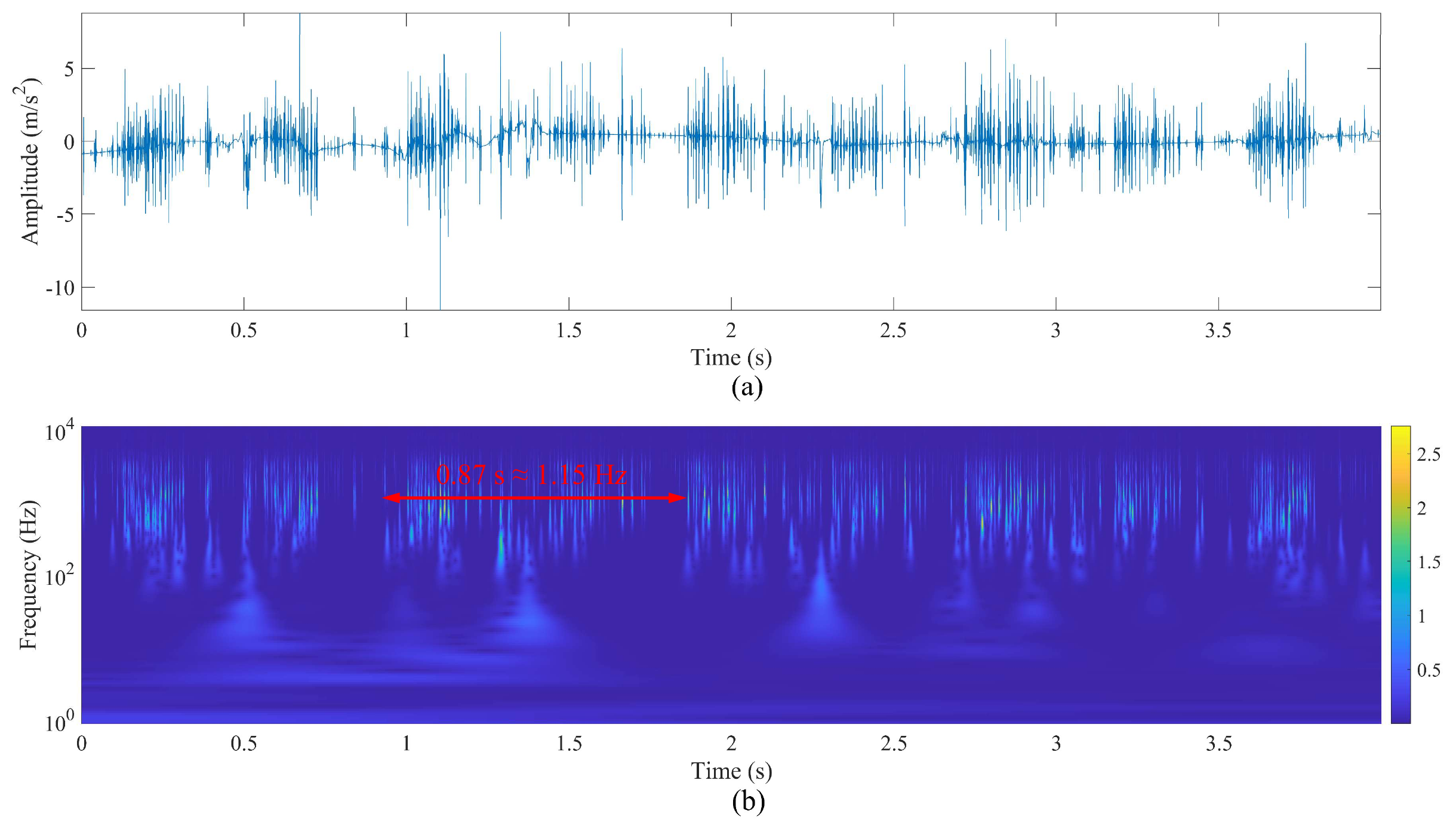}
	\caption{Time domain and frequency domain diagram of CGBBR. (a)time domain diagram of CGBBR, (b)Frequency domain diagram of CGBBR}
	\label{fig:time_frequecny}
\end{figure} 

\begin{figure}
    \centering
    \subfloat[IRRCC shares the same direction with ORRCC]
    {
      \includegraphics[width=0.8\linewidth]{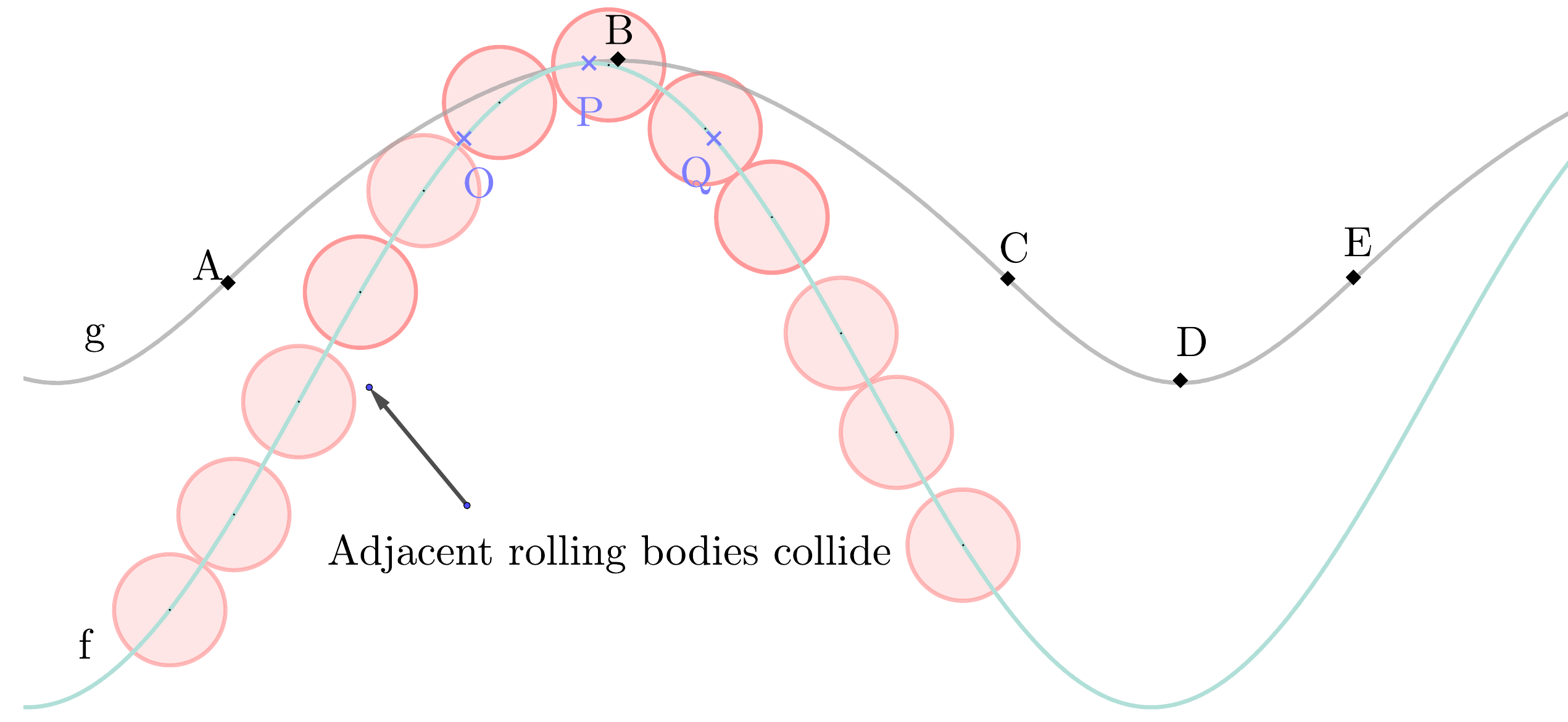}
      \label{fig:subfig1}
    }
    \quad
    \subfloat[IRRCC share the opposite direction with ORRCC]{
      \includegraphics[width=0.8\linewidth]{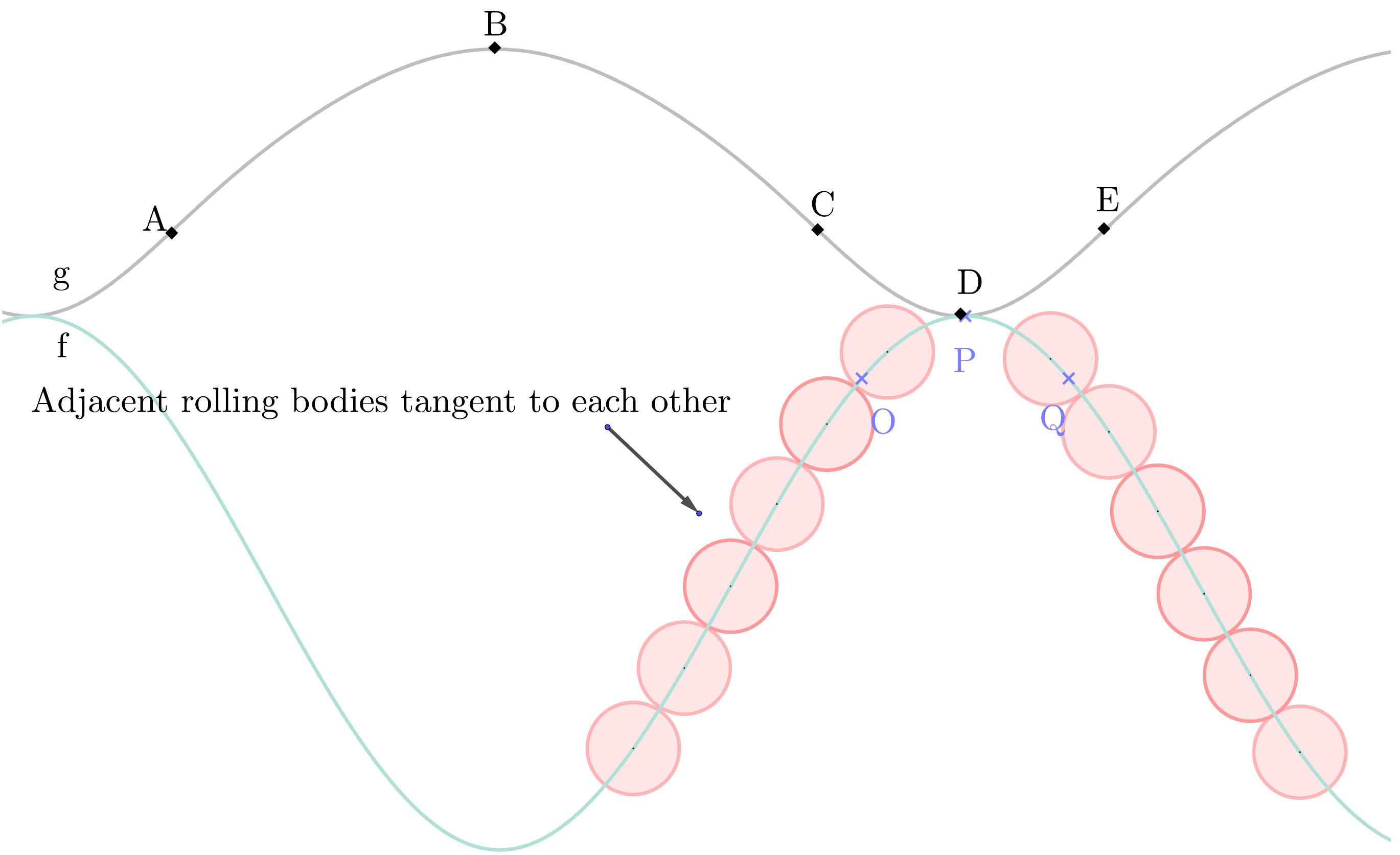}
      \label{fig:subfig2}
    }
    \caption{Influence of two contact states of IRRCC and ORRCC on rolling elements}
    \label{fig:kinematic vibration analysis}
\end{figure}

The vibration signal of CGBBR was detected, and its vibration acceleration data was 
successfully obtained. Fig. \ref{fig:time_frequecny} presents the time domain and 
time-frequency domain characteristics of the signal. The signal has a duration of 4 seconds 
and its shaft rotates at 1.15 Hz\cite{weiKinematicalAnalysesTransmission2011}. 
The analysis of both time-domain and time-frequency diagrams 
revealed periodic pulse clusters in the vibration signal of CGBBR. The time-domain and 
time-frequency diagrams exhibit that the vibration signal of CGBBR manifests periodic pulse 
clusters\cite{prudhomTimefrequencyVibrationAnalysis2017,chenVibrationSignalsAnalysis2020,randallWhyEMDSimilar2023}. 
This distinctive characteristic contrasts with ordinary bearings, which produce 
periodic pulses only when their raceways sustain damage. Therefore, the vibration 
characteristics of CGBBR differ significantly from those of conventional bearings.
The time-frequency diagram reveals that the periodic pulse clusters occur at the 
same frequency as the reciprocating motions within one revolution of the CGBBR. 
Moreover, it is worth noting that the duration of the periodic pulse clusters' appearance 
is considerably longer than that of their disappearance.

Based on the distinct attributes of CGBBR as a full complement roller bearing, it can be 
deduced that these pulse clusters originate from the stochastic interactions of rolling 
elements within the inner ring during the transition from ``valleys'' to ``peaks''.
To gain a clearer 
understanding of the mechanism behind the periodic pulse cluster, a schematic diagram is 
drawn to depict the contact state between rolling elements at both the ``valley'' and ``peak'' 
of CGBBR. 
As illustrated in Fig. \ref{fig:kinematic vibration analysis}, ORRCC 
functions as an envelope and remains tangent to the OPQ segment of the IRRCC. When the OPQ 
segment of IRRCC contacts the ABC segment of ORRCC, both curves share the 
same direction. This causes the rolling elements to shift towards the center of the OPQ 
segment, subsequently leading to a looser spacing between the rolling elements on either 
side of the OPQ segment. This results in collisions, which appear as pulse clusters in the 
time-frequency diagram. When the OPQ segment of the IRRCC approaches and contacts the CDE 
segment of the ORRCC, the opposite direction causes the rolling elements of the OPQ segment to 
be divided at the P point of the IRRCC. Consequently, rolling elements are pushed to both 
sides, and the adjacent rolling bodies become tangent without causing collisions. As a 
result, pulse clusters are not generated. The duration of the pulse clusters is longer when 
they appear compared to when they disappear, which can be attributed to the ABC segment being 
longer than the CDE segment.

\section{Conclusion}
\label{sec:conclusion}

CGBBR is a novel type of bearing with a unique structure, kinematics characteristics, and 
vibration response that distinguishes it from conventional bearings. This paper focused on 
the composition and design of CGBBR, as well as the motion laws and vibration responses of 
its components, which led to the following findings:
% \begin{enumerate}
%     \item Utilizing the envelope theory and swept surface theory, we successfully designed a curved groove ball bearing without a retainer(CGBBR), capable of converting rotary and reciprocating motions, as well as facilitating compound motion combining both. The effectiveness of this motion conversion mechanism was experimentally validated, demonstrating its potential in various applications.
%     \item we proposes the diameter-stroke ratio as a critical parameter for the subsequent optimization design of CGBBR. Furthermore, a method for calculating the number of rolling elements is introduced, providing an essential theoretical foundation for the further investigation of this bearing.
%     \item Through experimentation, we found that the vibration signals in CGBBR display periodic pulse clusters. We investigated the cause, determining that it arises when the central curves of the inner and outer raceways align, leading to collisions among the rolling elements. This pulse cluster phenomenon is exclusive to CGBBR and absent in traditional circular bearings.
% \end{enumerate}
\begin{enumerate}
    \item By utilizing the closed curve envelopment theory, a curved groove ball bearing without retainer (CGBBR) was successfully designed, capable of converting rotary and reciprocating motions, as well as facilitating compound motion combining both. The effectiveness of this motion conversion mechanism was experimentally validated, demonstrating its potential for various applications.
    \item The diameter-stroke ratio was proposed as a critical parameter for the subsequent optimization design of CGBBR. Additionally, a method for calculating the number of rolling elements was introduced, providing an essential theoretical foundation for further investigation of this bearing.
    \item It was discovered that the vibration signals in CGBBR exhibit periodic pulse clusters. The cause of this phenomenon was investigated, and it was determined that it arises when the central curves of the inner and outer raceways align, leading to collisions among the rolling elements. This pulse cluster phenomenon is exclusive to CGBBR and absent in traditional circular bearings. The findings were obtained through experimentation.
\end{enumerate}

Owing to the compact structure and sophisticated motion control capabilities of CGBBR, it 
holds extensive application prospects in various fields. For instance, it can substitute the 
crank-slider mechanism in engines for motion conversion, acting as the driving mechanism in 
compressors, or operating as a control component managing reciprocating and rotary motions in 
industrial robot joints. This paper lays the groundwork for design optimization, 
standardized production, and future broad implementation of curved groove ball bearing 
without retainer.

\section*{Declaration of interest}
\label{Declaration of interest}

The authors declare that they have no known competing financial interests or personal relationships that could have appeared to influence the work reported in this paper.

\section*{Data availability}
\label{Data availability}

Data will be made available on request.

\section*{Acknowledgement}
\label{Ackonwledgement}

This research was supported by the Talent start-up Project of Zhejiang A\&F University Scientific Research Development Foundation (2021LFR066).

%% The Appendices part is started with the command \appendix;
%% appendix sections are then done as normal sections
%% \appendix

%% \section{}
%% \label{}

%% For citations use: 
%%       \citet{<label>} ==> Jones et al. [21]
%%       \citep{<label>} ==> [21]
%%

%% If you have bibdatabase file and want bibtex to generate the
%% bibitems, please use
%%
\bibliographystyle{elsarticle-num-names} 
\bibliography{Manuscript.bib}

%% else use the following coding to input the bibitems directly in the
%% TeX file.

% \begin{thebibliography}{00}

%% \bibitem[Author(year)]{label}
%% Text of bibliographic item

% \bibitem[ ()]{}

% \end{thebibliography}
\end{document}